\documentclass{svjour2}                    
\smartqed  

\usepackage{graphicx}
\usepackage{amsmath,amssymb,bbm}
\usepackage{amsfonts}

\newcommand{\ket}[1]{\vert#1\rangle}
\newcommand{\xb}{\mathbf{x}}
\newcommand{\rb}{\mathbf{r}}
\newcommand{\sx}{\mathbf{s}}
\newcommand{\const}{\mathrm{const}}
\newcommand{\nucl}{\mathrm{nucl}}
\newcommand{\eq}{\mathrm{eq}}
\begin{document}

\title{Gravity-related wave function collapse}
\subtitle{Is superfluid He exceptional?}
\author{Lajos Di\'osi\thanks{The author thanks the organizers of the  International Workshop on Horizons of Quantum Physics for their invitation and generous support.}}

\institute{   Lajos Di\'osi\\
              Wigner Research Center for Physics, H-1525 Budapest 114, P.O.Box 49, Hungary\\
              Tel.: +36-30-3529648\\
              Fax: +36-1322-1710\\
              \email{diosi.lajos@wigner.mta.hu}
}

\date{Received: date / Accepted: date}

\maketitle

\begin{abstract}
The gravity-related model of spontaneous wave function collapse, 
a longtime hypothesis, damps the massive Schr\"odinger Cat states in quantum theory.
We extend the hypothesis and assume that spontaneous wave function 
collapses are responsible for the emergence of Newton interaction. 
Superfluid helium would then show significant and testable gravitational 
anomalies.
\end{abstract}

\section{Introduction}
\label{S1}
Quantum and gravity are expected to interfere in relativistic cosmic
phenomena only. Alternative speculations suggest that quantum and
gravity meet in a new way already at nanoscales. At least this 
follows from the gravity-related spontaneous collapse (decay) 
model \cite{Dio86,Dio87,Dio89,Pen94,Pen96,Pen98,Pen04} of massive macroscopic 
quantum superpositions also called Schr\"odinger cat (Cat) states. 
The DP model is non-relativistic, the predicted collapse rate is proportional 
to the Newton constant $G$, and it becomes relevant at nanoscales already. 
Although spontaneous collapses are too tiny, efforts are being under way to detect them, 
cf., e.g., \cite{Adl07,Mar_Bou03,Chr05,Van_Asp11,Rom11,LiKheRai11,Pep_Bou12}. 
  
I outline a way to go beyond the DP model. A self-contained and elementary explanation and
discussion of the DP model in Secs.~\ref{S2}-\ref{S5} is followed by a hypothesis 
extending the DP model. Sec.~\ref{S6} argues and conjectures that the Newton 
interaction is induced by the spontaneous collapses, 
it emerges stochastically at the same rate as the G-related collapse rate 
which is of the order of $1$ms for common condensed matter. 
However, for superfluid He the collapse time can be much longer!
Our most challenging conjecture is that the emergence time of gravity
depends on the quantum mechanical microstructure of the source. 
The Newton field of bulk superfluid He goes after the acceleration of the 
source at a substantial delay compared to common condensed matter (Sec.~\ref{S7}). 
Work \cite{Dio13} is closely related to Secs.~\ref{S2}-\ref{S5}, discusses 
the mass resolution and environmental decoherence issues in particular, 
and it conjectures the anomalous low collapse rate in liquid He. 

\section{Schr\"odinger Cat Problem}
\label{S2}
If we extend quantum mechanics from its original atom physical
context for larger objects, we face the well-known 
Cat paradox. To put it concrete and tractable, we often consider
mechanical Cat \cite{Dio86,Dio87,Dio89,Pen94,Pen96,Pen98,Pen04,Adl07}, 
e.g., a rigid massive ball in a superposition of two macroscopically different locations
$\xb$ and $\xb'$:
\begin{equation}
\ket{\mathrm{Cat}}=\ket{\xb}+\ket{\xb'}.
\label{Cat}
\end{equation}
This superposition is discontinued when a position measurement is 
applied to the ball, then it collapses randomly into $\ket{\xb}$
or $\ket{\xb'}$:
\begin{equation}
\ket{\mathrm{Cat}}\Longrightarrow\left\{\begin{array}{c}\ket{\xb}\\
                                        \ket{\xb'}\end{array}\right. .
\label{Coll}
\end{equation}
We applied standard quantum mechanics to the c.o.m. motion
of the ball as if it were a single atom or molecule. The Cat paradox lies
in that  we never see a massive body superposed at
two macroscopically different locations. The existence of the Cat 
state (\ref{Cat}) is  unlikely, its preparation seems impossible. 

However, the Cat state (\ref{Cat}) and collapse (\ref{Coll}) imply 
more than a paradox, they imply an acute physical problem. While in collapses 
the basic conservation laws are statistically respected,  a single branch 
after the collapse would violate them. Most embarrassing is the  
non-conservation of local mass density, i.e., violation of continuity. 
But non-conservation of total energy and momentum, the non-conservation 
of c.o.m. are also bad features. In our example,  the c.o.m. of the Cat state 
is $(\xb+\xb')/2$ before the collapse, and it becomes either 
$\xb$ or $\xb'$ randomly after the collapse. Note that breakdown of conservation
laws  in separate branches of the collapse is acceptable in quantum
measurement of a microscopic system because of its interaction with 
the measuring device. The same is not true for the Cat.
The measuring device can be almost massless compared to the Cat, 
the device won't be able to compensate the shift of the Cat c.o.m. 
caused by the collapse. 

The Cat paradox can be relaxed if one assumes that the collapse (\ref{Coll}) 
happens spontaneously and universally, in the absence and independent of any measurement 
devices. Accordingly,  one adds random spontaneous collapses to the 
standard quantum theory, resulting in non-linear and stochastic modification of 
the standard Schr\"odinger equation.  Such dynamical collapse models \cite{BasGhi03}
quickly damp the Cat states, ensuring that macroscopic variables have definite values, 
i.e., their quantum uncertainties remain suitably microscopical. 
But spontaneous universal collapse is no solution to
the violation of conservation laws. Dynamical collapse theories themselves would
steadily pump energy into the quantum system in question, this annoying side-effect 
has been known longtime ago, cf.~\cite{Pea00} and references therein. 

In the gravity-related spontaneous collapse model (Sec.~\ref{S3}) a plausible way out appears:
the energy-momentum balance might be restored by the gravitational field itself (Sec.~\ref{S6}).   

\section{Gravity-related spontaneous collapse}
\label{S3}
When constructing the mechanism of spontaneous collapse in
order to eliminate massive Cat states more general than (1),  
we assume that the spatial  mass density $f(\rb)$ of our quantum system 
matters. The Cat state reads   
\begin{equation}
\ket{\mathrm{Cat}}=\ket{f}+\ket{f'},
\label{Cat_f}
\end{equation}
where $f$ and $f'$ are 'macroscopically' different. This difference
---I call it 'catness'--- will be measured by a certain distance $\ell_G(f,f')$. 
The DP model \cite{Dio86,Dio87,Dio89,Pen94,Pen96,Pen98,Pen04} defines catness  by the following combination of 
three different Newtonian interaction potentials:
\begin{equation}
\ell_G^2(f,f')=2U(f,f')-U(f,f)-U(f',f')
\label{catness}
\end{equation}
where 
\begin{equation}
U(f,f')=-G\int\int\frac{f(\rb)f'(\sx)}{\vert\rb-\sx\vert}d\rb d\sx.
\label{Uff}
\end{equation} 
As we said, we assume that Nature makes $\ket{\mathrm{Cat}}$ decay spontaneously:
\begin{equation}
\ket{\mathrm{Cat}}\Longrightarrow\left\{\begin{array}{c}\ket{f}\\
                                        \ket{f'}\end{array}\right. .
\label{Coll_f}
\end{equation}
Small catness $\ell_G$ should allow for slow decay, large catness should spark fast decay.
Hence the DP model postulates the Cat characteristic life-time in the form $\tau_G=\hbar/\ell_G^2(f,f')$
which, with (\ref{catness}), yields the following collapse rate:
\begin{equation}
\frac{1}{\tau_G}=\frac{2U(f,f')-U(f,f)-U(f',f')}{\hbar}.
\label{rate}
\end{equation}
This expression guarantees immediate decay (fast rate) for macroscopic cats (Cats) and 
no decay (extreme slow rate) for atomic 'cats'. 

The DP model diverges for point-like constituents. If, e.g., we consider a single point-like
object of mass $M$ at location $\xb$ (and $\xb'$), it yields singular mass distributions: 
\begin{eqnarray}
f(\rb)&=&M\delta(\rb-\xb),\nonumber\\
f'(\rb)&=&M\delta(\rb-\xb').
\label{f_pointlike}
\end{eqnarray}
The  self-interaction terms $U(f,f)$ and   $U(f',f')$ become $-\infty$,
leading to a divergent collapse rate (\ref{rate}). 

\section{Resolution of mass density $f(\rb)$ matters}
\label{S4}
To treat the divergence of the DP model, we introduce a short length cutoff 
to limit the spatial resolution of the mass density $f$, i.e., to coarse-grain $f$. 
Two extreme cutoffs have been considered: i) $f$ is resolved microscopically down to the nuclear
size $\sim\!\!10^{-12}$cm or ii) f is coarse-grained over the atomic scales $\sim\!\!10^{-8}$cm. 
As we show below, the
predicted collapse rates are extremely different: 
for the microscopic resolution ($10^{-12}$cm) collapse may take milliseconds, 
for the macroscopic resolution ($10^{-8}$cm) it may take hours.
To see all this, we return to our mechanical Cat (\ref{Cat}) 
and apply the DP model (Sec.~\ref{S3}) to it. 

We consider the c.o.m. motion of a rigid spherical object
of macroscopic mass $M$ and radius $R$. Let us first assume that the
spatial cutoff is the bigger one ($10^{-8}$cm), 
then we coarse-grain $f$ (and $f'$) over the atomic
structure. E.g., a macroscopically homogeneous ball yields
\begin{eqnarray}
f(\rb)&=&\rho\theta(\vert\rb-\xb\vert\leq R),\nonumber\\
f'(\rb)&=&\rho\theta(\vert\rb-\xb'\vert\leq R),
\label{f_R}
\end{eqnarray}
where $\rho=M/(4\pi R^3/3)$ is the constant mass density, $\theta$ is the step-function.
The central quantity 
is the collapse rate (\ref{rate}) of the c.o.m. wave function.
It becomes the function of  the c.o.m. 
distance $\Delta\xb=\xb-\xb'$, and we calculate it in the first non-vanishing order \cite{Dio86,Dio87,Dio89}: 
\begin{equation}
\frac{1}{\tau_G}=\const\times\frac{M\omega_G^2}{\hbar}(\Delta\xb)^2.
\label{rate_macro}
\end{equation}
The $R$-dependence has been absorbed into the parameter $\omega_G=\sqrt{4\pi G\rho/3}$
which we call the frequency of the Newton oscillator, cf. Appendix.
This frequency is purely classical, its order of magnitude is $\omega_G\!\!\sim\!\!10^{-3}$/s
in typical condensed matter where $\rho$ is a few times $1$g/cm$^3$. 
The above collapse rate expression is valid if $\vert\Delta\xb\vert\ll R$. 

Alternatively, we consider the DP model with the 'nuclear' cutoff $10^{-12}$cm.
We therefore take $f$ such that the total mass $M$ is localized in the nuclei of size $\sim\!\!10^{-12}$cm 
and of density $\rho^\nucl\!\!\sim\!\!10^{12}\!\!\times\!\!\rho$.
To recalculate the collapse rate (\ref{rate}) for very small displacements $\vert\Delta\xb\vert\ll\sigma$, 
note that we can ignore pair-wise contributions of the nuclei.
Each nucleus contributes separately like a single ball of density $\rho^\nucl$.
Hence their total contribution amounts to the expression (\ref{rate_macro}) with $\omega_G^\nucl$
instead of $\omega_G$: 
\begin{equation}
\frac{1}{\tau_G}=\const\times\frac{M(\omega_G^\nucl)^2}{\hbar}(\Delta\xb)^2,
\label{rate_micro}
\end{equation}
where $\omega_G^\nucl=\sqrt{4\pi G\rho^\nucl/3}$ is the frequency of the Newton oscillator in nuclear matter.  
Its order of magnitude is  $\omega_G^\nucl\!\!\sim\!\!10^3$/s, cf. Appendix.
The rate expression (\ref{rate_micro}) is valid if $\vert\Delta\xb\vert\ll10^{-12}$cm. 

With the 'nuclear' resolution, Cat life-time $\tau_G$ has become cca $10^{12}$ times shorter! 
Without this huge enhancement \cite{Dio07} of the spontaneous collapses, the experimental test of the DP model 
would be too requesting, as recognized in \cite{Rom11,LiKheRai11,Pep_Bou12}.

\section{Equilibrium rate of spontaneous collapse} 
\label{S5}
We are going to investigate a possible universal feature of the standard quantum dynamics
modified by the DP spontaneous collapses. We expect that the unitary and 
collapse mechanisms, respectively, reach a certain balanced coexistence. 
For the ideal case of free Cat motion an exact proof is known, cf. \cite{Dio89}. 
Here we present the underlying idea leading to the correct qualitative results \cite{Dio87}. 
For the free c.o.m. motion of the Cat, the standard kinetic term in the Schr\"odinger equation 
tends to spread the wave function, competing with the DP spontaneous collapses
which tend to shrink the wave function. These counteracting tendencies reach the 
balance (equilibrium) when the corresponding two rates are equal:
\begin{equation}
\frac{\hbar}{M(\Delta\xb)^2}\sim\frac{M\omega_G^2(\Delta\xb)^2}{\hbar},
\label{balance}
\end{equation}
where the l.h.s. is the rate of kinetic changes, the r.h.s. is the rate (\ref{rate_macro}) 
of collapses. Now, calculate the geometric mean of the l.h.s. and the r.h.s., you
get $\omega_G$, hence both l.h.s. and r.h.s. must be of the order of $\omega_G$! 

For us, the important conclusion is the following. 
The estimated equilibrium collapse rate of the free Cat is $\omega_G$ which is of
the order of $10^{-3}$/s:
\begin{equation}
\frac{1}{\tau_G^\eq}\sim\omega_G\sim\frac{1}{\mathrm{hour}}.
\label{rate_eq_macro}
\end{equation}
The equilibrium collapse rate is independent of the mass $M$ and size $R$
of the Cat. The equilibrium width of localization comes out as
$\vert\Delta\xb\vert\!\!\sim\!\!\sqrt{\hbar/M \omega_G}$ --- a tiny scale for massive Cats.
The equilibrium collapse rate $\omega_G$ is fully classical, independent of $\hbar$.   
However, the obtained rate, i.e.: one collapse per hour, seems too low to be 
relevant under natural circumstances.  

The situation turns around if we assume the DP model with the microscopic mass density
resolution: we expect enhanced equilibrium collapse rate.  
Indeed, $\omega_G$ in the balance condition (\ref{balance}) must be replaced
by $\omega_G^\nucl$, meaning that the equilibrium collapse rate is cca $10^6$ times
higher than before:
\begin{equation}
\frac{1}{\tau_G^\eq}\sim\omega_G^\nucl\sim\frac{1}{\mathrm{ms}}.
\label{rate_eq_micro}
\end{equation}
This is a remarkably high rate of spontaneous collapse.  Although it was obtained
for the free rigid ball Cat, it is plausible to expect that a similar universal equilibrium
collapse rate exist in long wavelength hydrodynamic modes of condensed (or just bulk) matter
under natural circumstances. 


We make a tactical decision and define the DP model with the nuclear size $10^{-12}$cm cutoff. 
Hence the effect of DP collapses are enhanced and likely to be relevant even under natural circumstances.
The decoherence caused by spontaneous collapses is the main prediction of the DP model 
(as well as of all dynamical collapse models \cite{BasGhi03}).
Unfortunately, the direct observation of the decoherence in equilibrium Cat is rather hopeless,
Therefore, as argued in \cite{Dio07} (cf. also in \cite{DioLuk87,Dio92}) the DP model should be
refined, developed, extended. The upgraded model should predict
more characteristic phenomena than the spontaneous collapses. 

\section{If G-related collapse is the cause of gravity?}
\label{S6}
Although the Newton interaction and/or the Einstein equation played a role in the arguments 
\cite{Dio86,Dio87,Dio89,Pen94,Pen96,Pen98,Pen04} leading to the DP model,  
the parameter G in the resulting DP model controls the spontaneous collapses instead of the 
Newton interaction. The G-related spontaneous collapses damp Cat states 
but there is no gravitational dynamics (Newton acceleration). We can introduce it by hand, through
the Newtonian pair-potential. But we choose the alternative option of an emergent Newton
interaction. 

The main motivation comes from what we claimed to be the Cat problem in Sec.~\ref{S1}.
The standard collapse of a massive Cat strongly violates conservation laws. This is
invariably so with spontaneous collapses. In the equilibrium situation of Sec.~\ref{S5},
the Cat state is being continuously damped by the spontaneous collapses which are
still violating the conservation laws --- on a tinier scale this time. One would think
that these stochastic defects of conservation might be theoretically corrected if we 
construct a stochastic external field to compensate them. Obviously, external fields cannot restore 
all conservation laws. The non-conservation of the local density could only be compensated 
topologically, not dynamically. But momentum non-conservation can, in principle, be cured 
dynamically. We have recently presented a combination
of analytic and heuristic arguments to show that a random external field to compensate
the statistical non-conservation of the Cat momentum will on average contribute to what 
the Cat's own Newton field should be \cite{Dio09}. Until a more rigorous, convincing proof 
becomes available we explore a different perspective. 
In a sense, we go one step back, we don't apply analytical tools
(like the Ito-calculus in \cite{Dio09}). We conjecture heuristically instead.   
  
We have argued previously why the G-related collapses might induce the phenomenon of
Newton field and interaction. We don't intend to concretize the mechanism of the emergence, 
rather we formulate the following plausible feature of it.    
\emph{If the G-related collapses induce Newton gravity then, independently of the
detailed mechanism, the emergence rate/time of Newton gravity is related (proportional) 
to the wave function collapse rate/time of the sources.} We outline how this hypothesis
would lead to a particular anomaly w.r.t. standard Newton theory. 

\section{Testing gravity's laziness: Is He exceptional?}
\label{S7}
Recall Sec.~\ref{S6} where we defined the DP model with the low (nuclear size $10^{-12}$cm) 
cutoff, leading to the equilibrium collapse time $\tau_G^\eq\!\!\sim\!\!1$ms. 
In Sec.~\ref{S6} we assumed that this time-scale should be the characteristic emergence 
time of Newton gravity.

Ref.~\cite{Dio12} has tried to reconciliate the existence of a $1$ms emergence time 
with state-of-the-art Newtonian gravity. 
The available experimental, both astronomic and laboratory, evidences 
have poor time-resolution, perhaps not better than $1$ms.    
How immediate is the creation of the Newton field of accelerating mass sources?
Apparently, we cannot exclude a delay if it is $\sim\!\!1$ms or less. This poor
state of art may be advanced if we re-analyze previous evidences (or perform new 
Cavendish experiments) with accelerating (revolving) sources. 
That is a decent task for classical Newton gravity research. If there is a delay, 
it may depend on the geometry (i.e.: on the wavelength) of the sources. 
However, it may be quite hard to propose a concrete non-relativistic model where 
the Newton interaction is delayed. We don't open this chapter now, neither did Ref.~\cite{Dio12}.

Rather we anticipate a phenomenon which should, in the postulated framework, have
experimental significance. Let us recall that the 'short' equilibrium collapse 
(and emergence) time $\tau_G^\eq\!\!\sim\!\!1$ms comes from the fine-grained granular
 subatomic mass distribution of the sources. This is so with typical condensed matter sources. 
Superfluid He is the only non-relativistic exception where separate nuclei have no identity,
no localization. A 'ball' of superfluid He, with given c.o.m., is rather described by a 
smooth mass density (\ref{f_R}) than the the granular one. Hence, in superfluid He, 
DP-collapses may happen very slowly, the Newton field would emerge with a very long delay. 
For ideal homogeneous Cat the equilibrium collapse (and Newton field emergence) time would be 
$1$h (\ref{rate_eq_macro}). Superfluid's mass density $f(\rb)$, when the c.o.m. is fixed, 
looks perhaps half-way between nuclear granularity and homogeneity,
standard condensed matter physics must yield the concrete answer. Perhaps,
the DP equilibrium collapse (and emergence) rate happens to be somewhere half-way between $1$ms 
and $1$h, say, $1$s. That would imply that using a liquid He source in a Cavendish experiment, 
the pendulum would not react for about a second if we accelerate (e.g. remove) the source. 

\section{Summary}
\label{S8}
We pointed out that the Cat quantum state represents an acute theoretical problem:
violation of elementary conservation laws, yielding, e.g., non-conservation of the momentum.
We have, in simple terms, recapitulated the basics of the DP model of gravity-related  
spontaneous wave function collapses. The DP model would successfully damp the
Cat states whereas it cannot treat the violation of conservation laws. 
On the other hand, DP collapses show a universal feature when they reach
a balance (equilibrium) with the counteracting unitary dynamics. 
The collapse time-scale turns out to be the fully classical 
quantity $1/\sqrt{4\pi G\rho^\nucl/3}\!\!\sim\!\!1$ms for condensed matter.
This circumstance makes us formulate a conjecture beyond the DP model. 
We argue that the equilibrium stochastic DP collapses of massive objects are 
accompanied by the stochastic emergence of the Newton field around these
objects, with the tendency to restore the momentum balance. 
Since DP collapses are now claimed to be responsible for the emergence of the
Newton interaction, the emergence time of gravity should 
be proportional to the collapse time $1$ms of the 
condensed matter source. It follows from our underlying arguments
that gravity's emergence time may depend on the quantum mechanical
structure. Superfluid He is exceptional, having a very long 
equilibrium collapse time (like a second, maybe). We predict that
a Cavendish experiment with superfluid He source would detect the
time-delay of the field if, e.g., we quickly remove the source.

The author has been aspiring after a consistent realization of 
\emph{gravity from collapse} by some suitable mathematical extension
of the DP collapse model. These aspirations which have brought limited 
successes so far, have surfaced the preliminary conjectures presented above. 
They are extremely speculative, more than the DP model itself. 
Yet, they can be falsified or proved in straightforward experiments.

\section*{Appendix: Newton oscillator}
Take a homogeneous ball of mass density $\rho$, bore a narrow diagonal hole through it, 
and gently place a probe somewhere into the hole. The probe oscillates harmonically
at frequency 
$$
\omega_G=\sqrt{4\pi G\rho/3}
$$ 
where $G$ is the Newton constant. 
It is remarkable that the frequency is the function of density $\rho$ only, 
it does not depend separately on the size and the mass of the ball. In typical
condensed matter the density $\rho$ is a few times $1$g/cm$^3$, the frequency of the 
Newton oscillator is $\omega_G\!\!\sim\!\!10^{-3}$/s, the period is as long as cca $1$h. 

Formally, we can consider the Newton oscillator inside a homogeneous ball of nuclear 
density 
$\rho^\nucl\!\!\sim\!\!10^{12}$g/cm$^3$. The oscillator frequency 
$$
\omega_G^\nucl=\sqrt{4\pi G\rho^\nucl/3}
$$ 
becomes of the order of $10^{3}$/s, the period is as small as cca $1$ms.   
\begin{figure}
\begin{center}
\includegraphics[width=.3\textwidth]{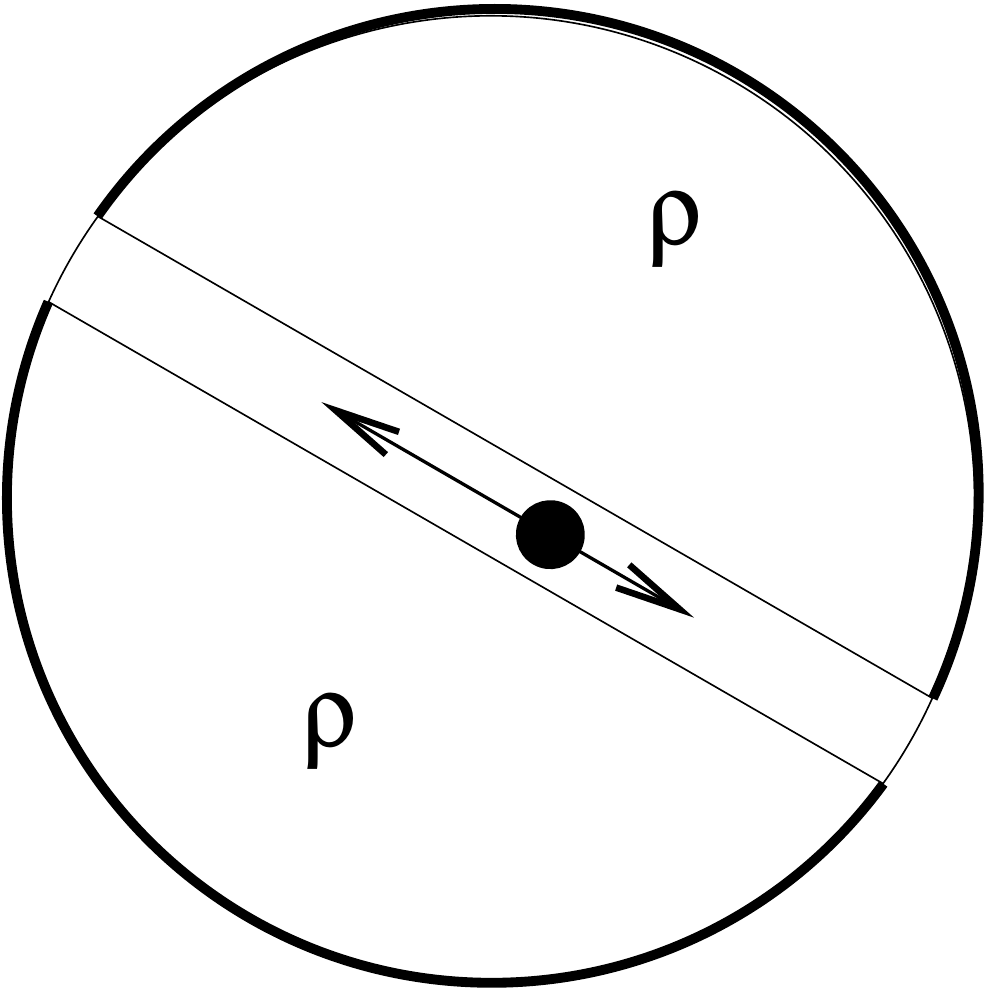}
\caption{Schematic view of a homogeneous ball of density $\rho$, with an infinite
narrow diagonal hole where the probe is oscillating at frequency $\omega_G=\sqrt{4\pi G\rho/3}$
under the directional force of the Newton field of the ball.}
\label{Fig}
\end{center}
\end{figure}

\begin{acknowledgements}
Support by the Hungarian Scientific Research Fund under Grant No. 75129
and support by the EU COST Action MP100 are acknowledged.
\end{acknowledgements}

\end{document}